\documentclass[fleqn,usenatbib]{mnras}

\usepackage{newtxtext,newtxmath}

\usepackage[T1]{fontenc}

\DeclareRobustCommand{\VAN}[3]{#2}
\let\VANthebibliography\thebibliography
\def\thebibliography{\DeclareRobustCommand{\VAN}[3]{##3}\VANthebibliography}

\usepackage{graphicx}	
\usepackage{amsmath}	

\newcommand{\be}{\begin{equation}}
\newcommand{\ee}{\end{equation}}

\newcommand{\Msun}{M_\odot}

\newcommand{\Mdotstar}{\dot{M}_\ast}
\newcommand{\Mdotin}{\dot{M}_\mathrm{in}}

\newcommand{\Pdot}{\dot{P}}

\newcommand{\rout}{r_\mathrm{out}}
\newcommand{\rin}{r_\mathrm{in}}
\newcommand{\rlc}{r_\mathrm{LC}}
\newcommand{\rco}{r_\mathrm{co}}
\newcommand{\rA}{r_\mathrm{A}}

\newcommand{\Ostar}{\Omega_\ast}

\newcommand{\Lacc}{L_\mathrm{acc}}

\newcommand{\Lcool}{L_\mathrm{cool}}
\newcommand{\Md}{M_\mathrm{d}}

\newcommand{\Lx}{L_\mathrm{X}}
\newcommand{\Lradio}{L_\mathrm{radio}}

\newcommand{\Tp}{T_\mathrm{P}}

\newcommand{\cs}{c_\mathrm{s}}

\newcommand{\Gammaacc}{\Gamma_\mathrm{acc}}
\newcommand{\GammaD}{\Gamma_\mathrm{D}}

\newcommand{\ergpers}{erg~s$^{-1}$}

\newcommand{\spers}{s~s$^{-1}$}

\newcommand{\Alfven}{Alfv$\acute{\mathrm{e}}$n~}
\newcommand{\srcb}{GLEAM-X J162759.5--523504.3}
\newcommand{\src}{PSR J0901--4046}

\usepackage{bm}

\title[Evolution of the long-$P$ pulsar \src]{Evolution of the long-period pulsar \src}

\author[A. A. Gen\c{c}ali et al.]{
A. A. Gen\c{c}ali,$^{1}$\thanks{E-mail: gencali@sabanciuniv.edu}
\"{U}. Ertan,$^{1}$
and M. A. Alpar$^{1}$
\\
$^{1}$Sabanc{\i} University, Orhanl{\i}, Tuzla, 34956, \.{I}stanbul, Turkey
}

\date{Accepted XXX. Received YYY; in original form ZZZ}

\pubyear{2022}

\begin{document}
\label{firstpage}
\pagerange{\pageref{firstpage}--\pageref{lastpage}}
\maketitle

\begin{abstract}
The fallback disc model predicted that anomalous X-ray pulsars (AXPs) and soft-gamma repeaters (SGRs) will evolve to isolated long period pulsars before the discovery of the first two long-period pulsars (LPPs) this year. Unlike normal radio pulsars, LPPs show transient pulsed-radio epochs with unusual and variable pulse shapes, similar to the radio behaviour of the few radio emitting AXP/SGRs. We show that the present properties of the recently discovered second LPP, \src~($P \simeq 76$~s), are obtained as a result of evolution in interaction with a fallback disc, as we had already shown for the first discovered LPP, \srcb~($P \simeq 1091$~s). While there is only an upper limit to the period derivative, $\Pdot$, of \srcb, the $\Pdot$ of the \src~has already been measured, providing better constraints for the evolutionary models. The model can produce the source properties with a dipole moment $\mu \simeq 10^{30}$ G cm$^3$. The results are not sensitive to the initial pulsar period. Our results indicate that \src~went through an AXP/SGR epoch at an age of a few $10^4$ yr, and is $\sim (6 - 8) \times 10^5$ yr old at present. 
\end{abstract}




\section{Introduction}
\label{intro}

All known radio pulsars and sources in isolated neutron star populations (i.e. not including neutron stars in binaries), namely anomalous X-ray pulsars and soft gamma repeaters (AXP/SGRs; also known as magnetars), dim isolated neutron stars (XDINs), high-magnetic-field radio pulsars (HBRPs), central compact objects (CCOs), and rotating radio transients (RRATs) have rotational periods smaller than $\sim 20$~s \citep[for a review see][]{Harding2013, Kaspi2016, Kaspi2017, Enoto2019}. An ultra-long-period pulsar, \srcb~(hereafter GLEAM-X), was discovered recently, with a period $P \simeq 1091$~s \citep{Hurley-Walker+2022}. This poses a challenge for long-term evolution models of isolated neutron stars spinning down under dipole radiation torques. The presence of torques from a fallback disc can explain the evolution and properties of such long-$P$ pulsars (LPPs), as had been predicted in the framework of the fallback disc model by \citet[][see fig. 3 and 4]{Benli2016} before the discovery of GLEAM-X. This work showed, through detailed simulations, that a neutron star evolving with a fallback disc and a dipole field strength of a few $10^{12}$~G would achieve $\sim 10^3$~s periods. Eventual descendants of AXP/SGRs evolving under torques from a fallback disc would indeed be such long-period sources with low X-ray luminosity, $\Lx$. The period $P \simeq 1091$~s of the first discovered LPP source GLEAM-X was reproduced by \citet{Gencali_2022} from evolutionary calculations with a fallback disc, satisfying the current upper limits on $\Lx$ and period derivative, $\Pdot$. 

A second LPP ($P = 75.89$~s), \src~(hereafter J0901), was discovered on September 27, 2020 with Meer(more) TRAnsients and Pulsars (MeerTRAP)\footnote{\url{ https://www.meertrap.org/}} and ThunderKAT\footnote{\url{http://www.thunderkat.uct.ac.za}} projects at the MeerKAT radio telescope in South Africa. For J0901, $\Pdot$ was estimated to be $2.25 \times 10^{-13}$~\spers, which corresponds to a rotational power, $\dot{E} = 4 \upi^2 I \Pdot / P^3 = 2 \times 10^{28}$~\ergpers, where $I$ is the moment of inertia of the neutron star, comparable to the observed pulsed radio luminosity $\Lradio$. The existence of a $\Pdot$ measurement makes this second source, \;J0901, more constraining for evolutionary models than GLEAM-X, for which only an upper limit for $\Pdot$ was estimated \citep{Hurley-Walker+2022}. The estimated distances for J0901 are $328$~pc and $467$~pc corresponding to the $\Lx$ upper limits of $\sim 1.6 \times 10^{30}$~\ergpers~and $\sim 3.2 \times 10^{30}$~\ergpers~respectively \citep{Caleb+2022}. From these $\Lx$ upper limits, the cooling age of J0901 is estimated to be greater than $\sim {\rm a~few} \times 10^5$~yr \citep[e.g.][]{Page2006, Page+2009, Potekhin2018, Potekhin2020}. A partially visible, diffuse, shell-like structure surrounding the source was observed, which could be the remnant of the supernova that formed the neutron star \citep{Caleb+2022}.

Both LPPs show variable pulse shapes during their transient radio-pulsar epochs. For comparison, five sources among the AXP/SGRs show radio pulsations \citep[][]{Kaspi2017, Esposito2020} in transient epochs after their X-ray outbursts \citep{Kaspi2017}. These sources also exhibit highly variable pulse profiles. Considering the similarities in their radio behaviour, \citet{Hurley-Walker+2022} suggested that the radio pulses of GLEAM-X could be generated by a mechanism similar to that operating in the radio emitting AXP/SGRs. 

Fallback discs around the new-born neutron star could form in supernova explosions \citep{Colgate1971, Michel1988, Chevalier1989, Perna2014}. It was proposed that neutron stars evolving with fallback discs can achieve the observed $\Lx$ and rotational properties of AXPs \citep{Chatterjee2000}. \citet{Alpar2001} proposed that including the properties of possible fallback discs as initial conditions in addition to the initial period and magnetic dipole moment could explain the emergence of the observed variety of single neutron star populations. Emission characteristics of fallback discs candidates were studied extensively for different sources \citep{Perna2014, Ertan2006, Ertan2007, Ertan2017, Ertan+2017, Posselt2018}. \citet{Ertan2007} showed that the emission from an irradiated disc can account for the observed broad-band spectrum of the AXP 4U 0142+61 from the optical to the mid-IR wavelengths \citep{Hulleman2000, Hulleman2004, Morii2005, Wang2006}. Further, it was shown through detailed calculations that the radiative processes produced by channelled mass accretion through the accretion column on to the neutron star can reproduce the observed soft and hard X-ray spectra, and the pulse profiles of AXP/SGRs \citep{Trumper2010, Trumper2013, Kylafis2014}. Long-term evolutionary models with fallback discs include the effects of the X-ray irradiation of the disc, the contribution of the cooling luminosity to the irradiation flux, and the eventual inactivation of the disc, which takes place when the temperature falls to a few $100$~K and thereby the ionisation fraction becomes so low that the magneto-rotational generation of viscosity and transport stops \citep{Balbus1991,Inutsuka2005}. These models were employed to understand the evolution and properties of individual sources from all isolated neutron star families \citep[see e.g.][]{Ertan2014, Benli2016, Gencali2021}.

In this work we investigate the long-term evolution for J0901 using the same fallback disc model applied earlier to GLEAM-X and to other isolated neutron stars from various categories. We show that the observed properties of J0901 can also be accounted for in this model. In Section~\ref{model}, we describe the model briefly. Our results are given in Section~\ref{results}, and the conclusions are summarised in Section~\ref{conc}.

\section{Model}
\label{model}

Here we briefly describe the disc parameters, initial conditions and evolution of a neutron star with a fallback disc. For details, the reader may refer to earlier work \citep[see e.g.][]{Ertan2014, Benli2016, Gencali2021}. We solve the disc diffusion equation starting with a steady state thin-disc profile using the kinematic viscosity $\nu = \alpha \cs h$, where $\alpha$ is the kinematic viscosity parameter, $\cs$ is the sound speed, and $h$ is the pressure scale height of the disc \citep{Shakura1973}. The three disc parameters governing the evolution are: (i) the irradiation efficiency parameter, $C$, (ii) the critical temperature, $\Tp$, below which the disc becomes viscously inactive, and (iii) the viscosity parameter $\alpha$. These parameters are expected to be similar for different systems evolving with fallback discs. Indeed, in all the earlier applications to different neutron star families, reasonable results were obtained with $\alpha = 0.045$, $C = (1 - 7) \times 10^{-4}$, and $\Tp = 50 - 150$~K \citep{Ertan2014, Benli2016, Benli2018, BenliCCO2018, Gencali2018, Ozcan2020, Gencali2021}. The initial conditions for a given source are the neutron star's magnetic dipole field strength, $B_0$ (on the poles), the initial rotational period $P_0$ of the star and the initial disc mass $\Md$. We keep $B_0$ constant during the evolution of a particular source, neglecting any magnetic field decay. 

Except for the very early stages of evolution, the sources always spin down throughout their lifetimes. In some epochs accretion takes place along with spin-down (ASD or weak propeller phase). In the ASD phase, we assume that all of the in-flowing disc matter is accreted on to the star from the co-rotation radius, $\rco = (G M / \Ostar^2)^{1/3}$, where $G$ is the gravitational constant, $M$ is the mass of the star, and $\Ostar$ is the angular speed of the star. In other epochs, when the source is in the strong propeller (SP) phase,  all of the disc inflow is thrown out from the inner disc and ejected out of the system. The transition between the ASD and SP phases takes place at a low $\Mdotin$ level as described below. There are three torque components acting on the star: (1) the magnetic spin-down torque, $\GammaD$, arising from the interaction between the inner disc and the magnetosphere of the star, (2) the spin-up torque, $\Gammaacc$, associated with accretion on to the star, and (3) the magnetic dipole spin-down torque, due to the electromagnetic radiation from the rotating dipole moment of the neutron star. The latter, which governs the spin-down of an isolated neutron star without a fallback disc, is usually negligible compared to the other components in the presence of a fallback disc. 

In the ASD phase, the inner disc radius $\rin = \rco$, and $\GammaD$ dominates $\Gammaacc$ unless the source is far from rotational equilibrium. $\GammaD$ is calculated by integrating the magnetic torques due to the disc, from the \Alfven radius, $\rA \simeq \big[ \umu^4/ (G M \Mdotin^{-2}) \big]^{1/7}$, to $\rco$ where $\umu$ is the magnetic dipole moment, and $\Mdotin$ is the mass-inflow rate of the disc. In terms of $\rA$ and $\Mdotin$, this calculation gives 
\begin{equation}
\label{disc_torque}
\GammaD = \frac{1}{2} \Mdotin~ \sqrt{G M \rA} 
\left[1- \left(\frac{\rA}{\rco}\right)^3\right] 
\end{equation}
\citep{ErtanE2008}.
Accretion exerts a spin-up torque $\Gammaacc = \Mdotstar \sqrt{G M \rin} $, on the star in the ASD phase. Here $\Mdotstar$ is the mass accretion rate on to the star, and we take $\Mdotstar = \Mdotin$ in the ASD phase. The X-ray luminosity $\Lx = \Lacc + \Lcool$ where $\Lacc = G M \Mdotstar / R_\ast$ is the accretion luminosity, $R_\ast$ is the star's radius and $\Lcool$ is the cooling luminosity of the neutron star, for which we use the theoretical cooling curve obtained by \citet{Page2006, Page+2009}. 

In the SP phase, $\Gammaacc = 0$, since $\Mdotstar = 0$. The transition from the ASD to the SP phase takes place at a low $\Mdotin$, usually when $\Lacc$ is below $\Lcool$. For $\Mdotin$ values that give $\rA$ larger than the light cylinder radius, $\rlc$, we replace $\rA$ with $\rlc$ in equation~(\ref{disc_torque}). We adopt the condition $\rA = \rlc$ for the ASD/SP transitions. This simplified transition condition was used in our earlier applications in the absence of a well known 
criterion for the onset of the SP phase that is consistent with observations. A more realistic model for the transition conditions and the torque calculations was developed recently \citep{Ertan2017, Ertan2018, Ertan2021}. This new analytical model predicts the $\Mdotin$ level for the onset of the propeller mechanism and the variation of $\rin$. This model will be used in the future for detailed analyses of evolutionary links between isolated neutron star families. In the present work, we use the simplified model for J0901 as well, for comparison with our earlier results. The more sophisticated new model \citep{Ertan2021} provides alternative paths for the past evolution but gives similar results for the present state and age of the LPP sources. 

The disc is heated through viscous dissipation and X-ray irradiation. For radii greater than a few $10^9$~cm X-ray irradiation dominates over viscous dissipation and significantly affects the active life-time of the disc. The outer radius of the viscously active disc, $\rout$, is the radius beyond which the effective temperature is currently below $\Tp$. During the long-term evolution, $\Lx$ and the irradiation strength decrease gradually, causing $\rout$ to propagate inwards. Eventually, the entire disc becomes inactive at an age smaller than $10^6$~yr. The disc viscosity drops, mass inflow and accretion stop and the star continues to spin down by the weak magnetic dipole torque alone. 

For a particular source, we trace various initial conditions through many simulations to check whether the model can produce the source properties. We then determine the allowed ranges of initial conditions that give model curves reproducing properties of the source at a reasonable present age. Our results for J0901 are summarised in the next section.

\section{Results \& Discussion}
\label{results}

\begin{figure}
    \centering
    \includegraphics[width=\columnwidth]{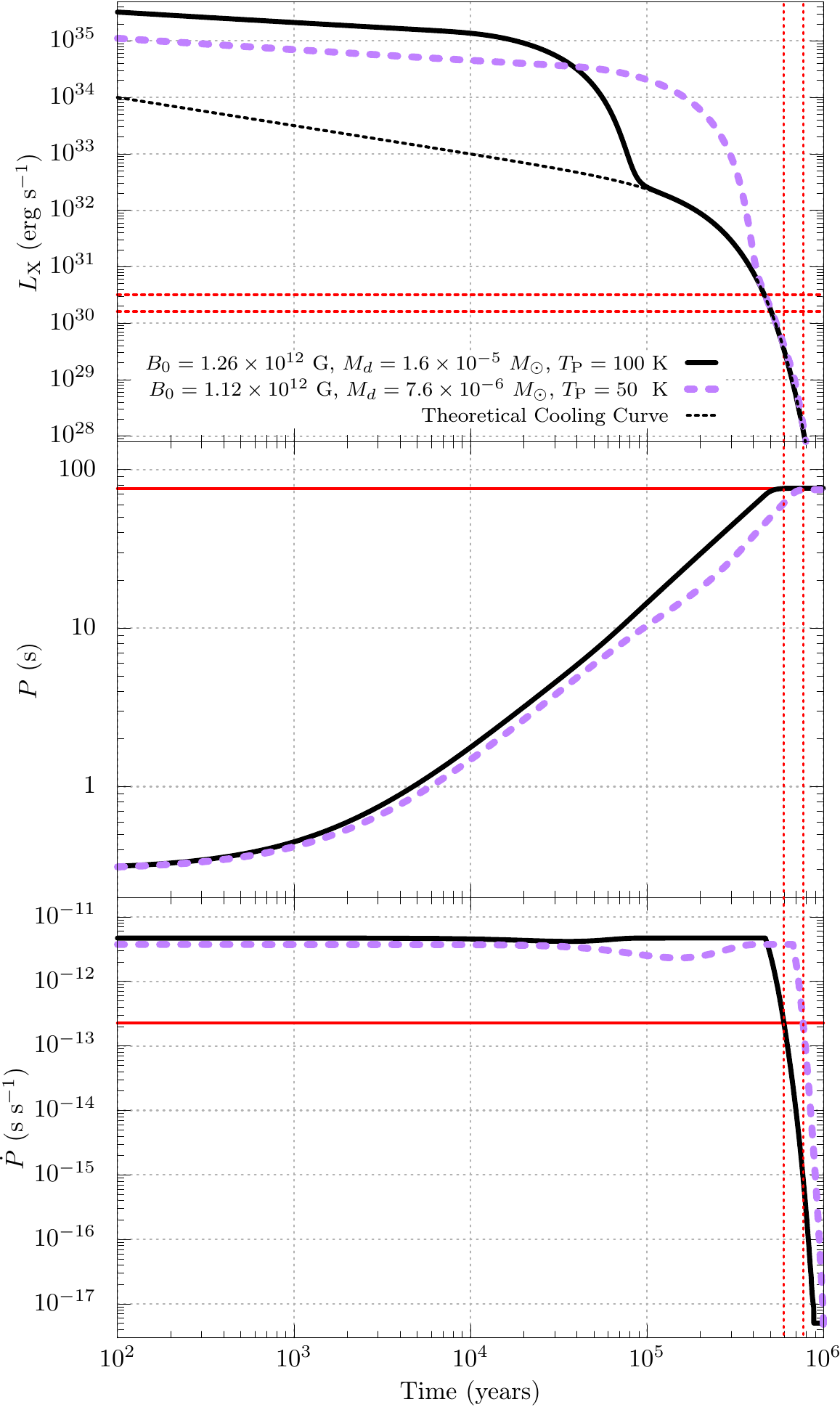}
    \caption{Illustrative model curves for the long-term evolution of J0901. These curves are obtained with the initial conditions $B_0$ and $\Md$ values given in the top panel, while $P_0 = 0.3$~s for both model curves. For both models, we take $\alpha = 0.045$, and $C = 7 \times 10^{-4}$. The $\Tp$ values are given in the figure. In the top panel, the horizontal dashed lines show the estimated the upper limits. The dotted curve shows the evolution of the cooling luminosity $\Lcool$ of the star \citep{Page2006, Page+2009}. The solid horizontal lines in the middle and bottom panels show the measured $P$ and $\Pdot$ of the source respectively. These illustrative models achieve the current rotational properties of the source at the ages shown by the vertical dotted lines.}
    \label{fig1}
\end{figure}

Illustrative model curves seen in Fig.~\ref{fig1} show that the model can reproduce the $P$ and $\Pdot$ of J0901 simultaneously with a range of $\Lx$ values above a minimum $\Lx \sim 10^{28}$~\ergpers~and consistent with the estimated $\Lx$ upper limits ($\sim 10^{30}$~\ergpers). The current properties are reached at an age $t \sim (6 - 8) \times 10^5$~yr, while the source is in the SP phase, slowing down with the $\GammaD$. These model curves are obtained with disc parameters $\alpha$, $C$ and $\Tp$ similar to those employed for GLEAM-X and other sources from different neutron star families evolving under fallback disc torques. Our results are not sensitive to the $P_0$. For the model curves in Fig.~\ref{fig1}, $\Md = 1.6 \times 10^{-5}~\Msun$ and $\Md = 7.6 \times 10^{-6}~\Msun$ for the solid and the dashed curves respectively, while $P_0 = 0.3$~s for both model curves. We obtain reasonable model curves for J0901 with $B_0 \sim 10^{12}$~G. This dipole field strength together with the $76$~s period place the source below the pulsar death line \citep{Chen1993}, as was the case for the current evolutionary phase of GLEAM-X in the same model \citep{Gencali_2022}.  

Transient radio epochs and variable pulse profiles of J0901 and GLEAM-X are similar to the pulsed radio behaviour observed from a few AXP/SGRs. This might imply that a common radio emission mechanism is operating in these systems \citep{Hurley-Walker+2022}, possibly similar to the origin of the brief radio bursts of RRATs, for which the occasional enhancement of the magnetic field flux in open field lines by means of disc-field interaction could be responsible \citep{Parfrey2016, Parfrey2017, Gencali2021}. Even if a source is below the relevant pulsar death line, inward propagation of the inner disc radius and subsequent opening of the closed field lines could provide the power required for the radio emission. The same effect could produce different radio behaviour in different accretion and rotation regimes, and might power the transient radio pulses of the LPPs as well \citep[for discussion see][]{Parfrey2016, Parfrey2017, Gencali2021, Tong2022}. Details of the  physical processes responsible for the  observed radio emission of these sources are beyond the scope of the present work.  

\citet{Ronchi2022} recently proposed that a neutron star evolving with a fallback disc and a {\em magnetar} dipole field can achieve the rotational properties of GLEAM-X and J0901 at young ages ($\sim 10^3 - 10^5$~yr). Their model is quite different from the  model employed in this work. They use a simplified analytical model for the $\Mdotin$ evolution \citep{Menou2001,Ertan2009} without including the heating by the X-ray irradiation of the disc produced by accretion luminosity and the cooling of the neutron star; and assumed that the disc becomes inactive (at different $\Mdotin$ values) when the model sources reach their presently observed periods. Furthermore, at the ages of the LPPs predicted in model ($< 10^5$~yr), the $\Lx$ values estimated from the theoretical cooling curves \citep{Page2006, Page+2009, Potekhin2018, Potekhin2020} remain above a few $10^{32}$~\ergpers~even without contribution of the magnetar field decay \citep[see][]{Vigano2013}. This $\Lx$ level is two orders of magnitude greater than $\sim 10^{30}$~\ergpers, the $\Lx$ upper limit estimated for J0901. For a crustal field stronger than $10^{13}$~G, $\Lx$ is estimated to decrease below the upper limit  at an age $> 1$~Myr. Thus, \citet{Ronchi2022} model with simplifying assumptions reproduces the rotational properties of these LPPs at the expense of inconsistency  with the $\Lx$ upper limits \citep{Rea2022}.

Is there any evolutionary link between the LPPs and other neutron star populations? In our model, evolutionary paths of the two  LPPs pass through the ranges of $\Lx$ and rotational properties of AXP/SGRs at ages of a few $10^4$ yrs, similar to the ages estimated for present day AXP/SGRs (see Fig.~\ref{fig1}). From the model curves, we estimate that AXP/SGRs with relatively strong dipole fields ($B_0 \sim 10^{13}$~G) evolve to the GLEAM-X properties \citep{Gencali_2022}, while an average AXP/SGR ($B_0 \sim$ a few $10^{12}$~G) is likely to become a source similar to J0901 at $t \sim (6 - 8) \times 10^5$~yrs, in a state close to the final inactivation of the entire disc. These evolutionary paths leading AXP/SGRs to long periods were already predicted in our earlier work \citep[][see fig. 2 and 3]{Benli2016}. In this model, AXP/SGRs are not likely to exhibit normal radio pulsar behaviour. Since most of these systems are estimated to be in the ASD phase at present, accretion on to the star prevents their normal pulsed radio emission. On the other hand, after the transition to the SP phase, the radio emission would not be hindered by accretion, but their rotational power is not sufficient at their current long periods. In this late phase of evolution with very low $\Lx$, detections in X-rays are not likely. Emission of the disc in the IR bands is also weak because of the currently weak X-ray irradiation. In other words, if the LPPs were not able to show their unusual transient radio emission, we probably would not be aware of their existence. Long period neutron stars as endpoints of evolution with fallback discs without transient activity must constitute an abundant population, as is also the case for descendants of isolated radio pulsars which have spun down below the pulsar death valley.  

These results imply that the persistent AXP/SGRs are likely to be progenitors of the LPP population. The birth rate of AXP/SGRs is estimated to be $\sim 3 $~kyr$^{-1}$ \citep{Keane2008}. Less than half of the known AXP/SGRs are persistent \citep[11 sources;][see Magnetar Catalogue]{Olausen2014}\footnote{\url{http://www.physics.mcgill.ca/~pulsar/magnetar/main.html}} and can achieve long periods. From the simulations, we estimate that the periods of LPPs could range from a few $10$~s to a few $10^3$~s for the $B_0$ range estimated for AXP/SGRs with the same model. We have estimated the ages of the two LPPs around $5 \times 10^5$~yr, while the AXP/SGR ages vary from  $\sim 10^3$~yr to a few $10^4$~yr. We therefore estimate that the number of LPP candidates could be an order of magnitude greater than that of AXP/SGRs. 

Our results imply that the X-ray luminosities of LPPs sharply decrease to below $\sim 10^{31}$~\ergpers~after ages of a few $10^5$~yr. This means that older LPPs are not likely to be detected in X-rays with present generation observatories, except for nearby sources, and that they  are not likely to show normal radio pulsar behaviour either. Considering the differences in the observed $P$ and estimated $B_0$ of the two LPPs, we estimate that a significant fraction of LPP candidates could show detectable pulsed radio-burst epochs. However, the physical mechanism activating pulsed-radio emission and the recurrence times of the radio epochs are not known yet, one is not able to make statistical predictions about radio detections. 

\section{Conclusions}
\label{conc}

We have shown that a neutron star evolving with a fallback disc and a conventional dipole field  ($B_0  \sim 10^{12}$~G) can acquire the rotational properties of J0901 at an age of $\sim (6 - 8) \times 10^5$~yr. Our results are not sensitive to the $P_0$. The expected $\Lx$ is below the estimated $\Lx$ upper limit of the source. Our model results indicate that, like GLEAM-X, J0901 was also an AXP/SGR in the early phase of its evolution: The LPPs are evolutionary descendants of AXP/SGRs. The model can reproduce the properties of these sources with the same disc parameters ($\alpha = 0.045$, $C = (1-7) \times 10^{-4}$, and $\Tp = 50 - 150$~K) as those employed for the other young neutron star populations in earlier work. All isolated neutron star populations including AXP/SGRs, XDINs, HBRPs, CCOs, RRATs and the newly discovered LPPs can be explained in the fallback disc model with similar parameters of disc physics. The differences between the evolutionary paths arise from different initial conditions, namely the initial disc mass, magnetic dipole moment, and the initial period. We will continue to investigate the evolutionary links between all these populations in more detail in future work.       

\section*{Acknowledgements}

We thank the anonymous referee for their detailed constructive and helpful criticism on our manuscript.
We acknowledge research support from Sabanc{\i} University, and from T\"{U}B\.{I}TAK (The Scientific and Technological Research Council of Turkey) through grant 120F329.

\section*{Data Availability}

No new data were analysed in support of this paper.



\bibliographystyle{mnras}
\bibliography{example} 








\bsp	
\label{lastpage}
\end{document}